\documentclass[12pt,preprint]{aastex}

\slugcomment{Submitted to Astrophysical Journal}

\shorttitle{Solar Flare Ar abundances}
\shortauthors{Sylwester et al.}

\begin{document}

\title{A solar spectroscopic absolute abundance of argon from RESIK}


\author{J. Sylwester\altaffilmark{1} and B. Sylwester\altaffilmark{1} }
\affil{Space Research Centre, Polish Academy of Sciences, 51-622, Kopernika~11, Wroc{\l}aw, Poland}
\email{js@cbk.pan.wroc.pl}

\and

\author{K. J. H. Phillips\altaffilmark{2}}
\affil{Mullard Space Science Laboratory, University College London, Holmbury St Mary, Dorking,
Surrey RH5 6NT, U.K.}
\email{kjhp@mssl.ucl.ac.uk}

\and

\author{V. D. Kuznetsov\altaffilmark{3} }
\affil{Institute of Terrestrial Magnetism and Radiowave Propagation (IZMIRAN), Troitsk, Moscow, Russia}
\email{kvd@izmiran.ru}

\begin{abstract}
Observations of He-like and H-like Ar (\ion{Ar}{17} and \ion{Ar}{18}) lines at 3.949~\AA\ and 3.733~\AA\ respectively with the RESIK X-ray spectrometer on the {\it CORONAS~F} spacecraft, together with temperatures and emission measures from the two channels of {\it GOES},  have been analyzed to obtain the abundance of Ar in flare plasmas in the solar corona. The line fluxes per unit emission measure show a temperature dependence like that predicted from theory, and lead to spectroscopically determined values for the absolute Ar abundance, $A({\rm Ar}) = 6.44 \pm 0.07$ (\ion{Ar}{17}) and $6.49 \pm 0.16$ (\ion{Ar}{18}) which are in agreement to within uncertainties. The weighted mean is $6.45 \pm 0.06$, which is between two recent compilations of the solar Ar abundance and suggest that the photospheric and coronal abundances of Ar are very similar.
\end{abstract}


\section{INTRODUCTION}\label{intro}

There has been much recent interest in the abundance of argon in the Sun and nearby objects \citep{lanz07,lod08,asp09}. The abundance of argon in the solar photosphere cannot be directly derived as the temperature is much lower than that needed to excite neutral or singly-ionized Ar lines, so proxies such as nearby B stars, \ion{H}{2} regions, planetary nebulae, or solar system sources such as the atmosphere of Jupiter must be used. It is presumed that the Sun collapsed out of material that is represented by such objects, so the element abundances derived are ``protosolar": allowance must, however, be made for a certain amount of heavy-element settling over the Sun's lifetime.  There are several emission lines of Ar ions in the extreme-ultraviolet and X-ray spectra of the quiet or active region solar corona or solar flares, and such lines have been used to give Ar abundances \citep{fel90,you97}. Commonly line flux ratios are used to obtain the ratios of abundances from spectroscopic measurements; thus \cite{fel90} obtained argon-to-magnesium abundance ratio of $0.15 \pm 0.05$ from measurements with the {\it Skylab} ultraviolet spectrometer S082A, while \cite{you97} obtained an argon-to-calcium abundance ratio in the range $0.55-1.1$, from both {\it Skylab} and optical measurements during an eclipse. Using published element abundance sets for Mg and Ca, \cite{lod08} in her review has given a grand average of values from various solar determinations as $A({\rm Ar}) = 6.50 \pm 0.10$. (Abundances here are expressed on a logarithmic scale with H $=12$ by a notation extensively used in recent literature, $A({\rm Ar})$.)  The review by \cite{asp09} gives $A({\rm Ar}) = 6.40 \pm 0.13$, using a revised O abundance, which updates a previous determination \citep{asp05} of 6.10 based on a single measurement from solar energetic particles. The possibility of fractionation through the ``FIP" (first ionization potential) effect is recognized, with the abundances of elements having low FIP ($\lesssim 10$~eV) being enhanced in the solar wind and corona by factors of $\sim 4$ and abundances of elements with high FIP ($\gtrsim 10$~eV) approximately the same as the proxies for the solar photosphere \citep{fel92b,flu99}: Ar is of interest because its FIP (15.8~eV) is larger than that of any other common element in the Sun apart from He and Ne.

The high-resolution X-ray crystal spectrometer RESIK (REntgenovsky Spektrometr s Izognutymi Kristalami; \cite{syl05}) on board the {\it CORONAS-F} spacecraft has observed X-ray lines in the 3.3--6.1~\AA\ range emitted by solar coronal plasmas. Line fluxes with diagnostic potential including the possibility of deriving element abundances have already been analyzed \citep{phi03,syl08a,syl08b,bsyl10,syl10,phi10}. The instrument operated between 2002 and 2003, and several thousand spectra were collected. It had significant improvements over previous spectrometers in that a background formed by fluorescence of the crystal material was eliminated for the two short-wavelength channels by judicious settings of the detector high-voltages and pulse height discriminators, and minimized to very small levels for the remaining two channels. The period from 2002 August to 2003 February was particularly appropriate for analysis as the settings were then fully optimized, and spectra were selected from this period for the analysis described here.

Here we use flux values of the \ion{Ar}{17} resonance ($w$) line at 3.949~\AA\ (actually the flux in the 3.94--4.01~\AA\ interval minus the flux in a neighboring portion of continuum) to derive the absolute abundance of Ar (Ar/H) by comparing with theoretical values taken from the {\sc chianti} database and code \citep{der97,der09}. As the sensitivity of RESIK is very high, the flux measurements have good statistical quality and are a considerable improvement over those from the {\it Solar Maximum Mission} Flat Crystal Spectrometer \citep{phi94}. The Ly-$\alpha$ lines of \ion{Ar}{18} form a single prominent feature at 3.733~\AA\ in high-temperature flare spectra observed in RESIK channel~1, and are also available for analysis in the same way. The observations and analysis are described in Section~2, and comparison of argon abundances from the \ion{Ar}{17} and \ion{Ar}{18} lines and with values obtained by other workers is described in Section~3.

\section{OBSERVATIONS}\label{Observations}

The spectra discussed here were obtained in channels 1 (3.40--3.80~\AA) and 2 (3.83--4.27\AA) of the RESIK instrument. X-rays from the Sun were diffracted by silicon crystals (Si 111, $2d = 6.27$~\AA) for both these channels. RESIK channel~2 spectra  include \ion{Ar}{17} emission lines consisting of three features called $w$ (transition $1s^2\,^1S_0 - 1s2p\,^1P_1$, 3.949~\AA), $x+y$ ($1s^2\,^1S_0 - 1s2p\,^3P_{2,1}$, 3.969~\AA), and $z$ ($1s^2\,^1S_0 - 1s2s\,^3S_1$, 3.994~\AA). The line flux ratio $G = (x+y+z)/w$ is slightly temperature-dependent.

The $G$ ratios from measured line fluxes in particular time intervals were plotted against single values of temperature and emission measure $\int_V N_e^2 dV$ ($N_e = $ electron density, $V =$ emitting volume) estimated from the ratio of emission in the two channels of {\it GOES} \citep{whi05}, and were found to be close to theoretical values based on data from {\sc chianti} \citep{syl08a}. Although a slight improvement to the fit was obtained by taking a two-temperature-component emission measure, in this work we take the original isothermal model as more suitable for analyzing the large number of spectra available. Channel~1 spectra  include \ion{K}{18} lines which were previously discussed \citep{syl10} and, for temperatures greater than about 10~MK, the \ion{Ar}{18} Ly-$\alpha$ line feature at 3.733~\AA.

Some 2795 spectra from a sample of 20 flares between 2002 August and 2003 February, having {\it GOES} importance ranging from C1 to X1, were selected for analysis, and an average of non-flaring points taken from \cite{bsyl10}. The spectra were collected over integration times that were inversely related to the X-ray activity level, from 2~s at the peaks of strong flares to 5~minutes for non-flaring periods. Details of the selected spectra are given by \cite{bsyl10} and \cite{phi10}. Calibration factors, converting photon count rates to absolute flux units, were applied to all the spectra from post-launch analysis \citep{syl05}.

\begin{figure}
\epsscale{.80}
\plotone{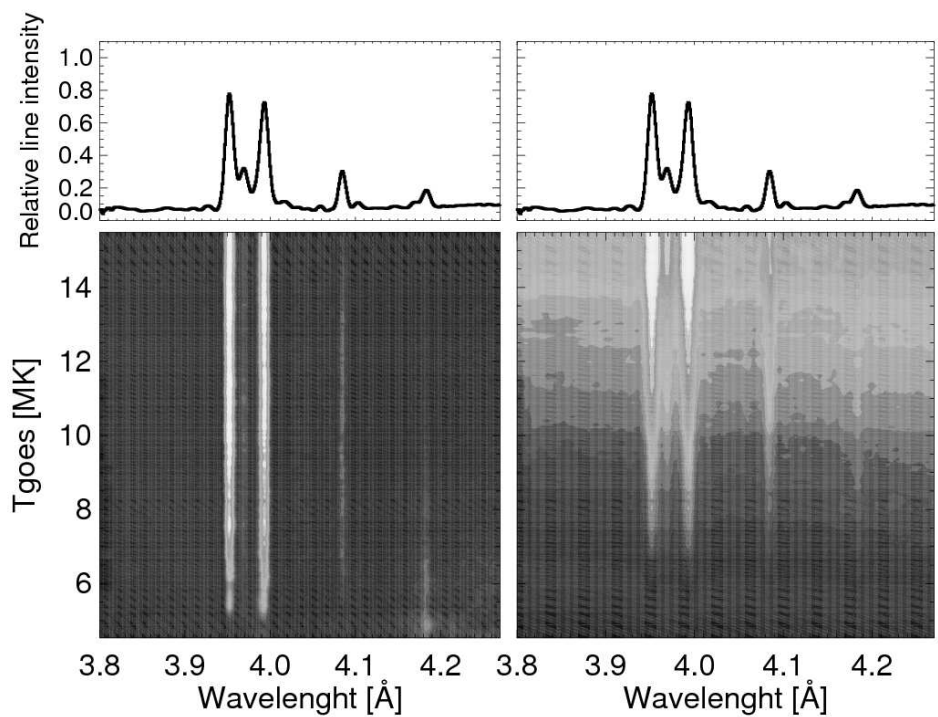}
\caption{RESIK channel~2 spectra stacked with increasing temperature $T_{\rm GOES}$ (in MK) in the vertical direction shown in a gray-scale representation (lighter areas indicate higher fluxes). The total spectrum is indicated at the top of each plot, with synthetic spectra (narrow line profiles) also shown. The principal lines are: \ion{Ar}{17} $w$, $x+y$, $z$ (3.95~\AA, 3.97~\AA, 3.99~\AA); \ion{S}{15} $w4$ (4.09~\AA), and the 4.19~\AA\ line feature, which is a blend of \ion{S}{14} satellites and the \ion{Cl}{17} Ly-$\alpha$ lines. {\it Left:} Not normalized. {\it Right:} Normalized to the total number of photon counts in each spectrum. } \label{stacked_RESIK_ch2_sp}
\end{figure}

The \ion{Ar}{17} lines in channel~2 were observed in spectra over a large temperature range. As in our previous work and in flare analyses by several other authors, we derived temperature ($T_{\rm GOES}$) from the flux ratio of the two {\it GOES} channels for the time period of each RESIK spectrum using the procedure of \cite{whi05}. Figure~\ref{stacked_RESIK_ch2_sp} (left-hand plot) shows all the spectra analyzed stacked vertically in order of $T_{\rm GOES}$. In the right-hand plot, the spectra are normalized to the total number of photon counts, and the total spectrum shown at the top of each plot. The \ion{Ar}{17} lines are evident over the entire temperature range shown, even at temperatures as low as 5~MK. As well as the principal \ion{Ar}{17} lines (3.949--3.994~\AA), the \ion{S}{15} $1s^2\,^1S_0 - 1s4p\,^1P_1$ line (also known as $w4$) is prominent at 4.088~\AA, as is a line feature at 4.19~\AA\ made up of \ion{S}{14} dielectronic satellites (transitions $1s^2\,2l - 1s2l4l'$, $l$, $l' = s$, $p$) and at higher temperatures ($T_e \gtrsim 10$~MK) the H-like Cl (\ion{Cl}{17}) Ly-$\alpha$ line.  Weak lines of \ion{S}{15} with transitions $1s^2-1snp$ with $n\eqslantless 10$ occur between 3.998~\AA\ ($w5$) and 3.883~\AA\ ($w10$) -- wavelengths are from atomic data extending the usual database of the {\sc chianti} database and code  \citep{der97,lan06,der09}.\footnotemark\

\footnotetext{Thanks are due to Dr E. Landi for supplying these data.}

The \ion{Ar}{18} Ly-$\alpha$ line at 3.733~\AA\ falls in the range of RESIK channel~1. It is emitted over a wide range of temperatures, though it is most prominent at temperatures above about 10~MK. The line is visible in channel~1 spectra already shown in \cite{syl10} in a stacked representation like Figure~\ref{stacked_RESIK_ch2_sp}. The \ion{Ar}{18} line feature occurs at very nearly twice the wavelength of the well known group of highly ionized Fe lines ($2 \times 1.85$~\AA) but these have no effect on the observed spectra as second-order diffraction is forbidden for Si 111 crystal used in channel~1. Weak \ion{S}{16} lines  ($1s-5p$, 3.696~\AA; $1s-4p$, 3.784~\AA) occur near the \ion{Ar}{18} line feature, but have no effect on the measurement of the \ion{Ar}{18} line flux.

\section{ANALYSIS AND RESULTS}\label{Results}

The fluxes of the \ion{Ar}{17} $w$ and \ion{Ar}{18} Ly-$\alpha$ lines were estimated by taking narrow wavelength intervals around the lines (3.94--4.01~\AA\ for the $w$ line and 3.72--3.745~\AA\ for the Ly-$\alpha$ line) and subtracting a portion of nearby continuum. Uncertainties in each flux measurement were taken from the uncertainties in the photon counts, assuming Poissonian statistics. For the \ion{Ar}{17} $w$ line, the continuum flux was invariably much less than the flux in the range containing the line, but this was not always the case for the \ion{Ar}{18} Ly-$\alpha$ line. We combined the uncertainties in the line and continuum flux measurements and if this combined uncertainty for any spectrum was larger than the difference between the line and continuum flux we rejected the estimate.

\begin{figure}
\epsscale{.80}
\plotone{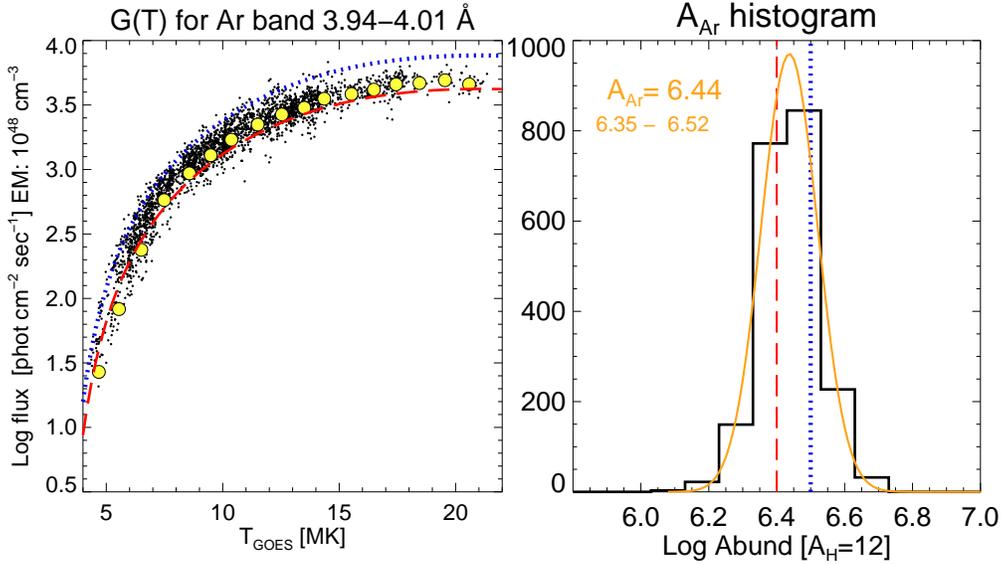}
\caption{(Left panel:) Estimated fluxes (small black points) in the \ion{Ar}{17} $w$ line divided by the {\it GOES} emission measure $EM_{\rm GOES}$ (units of $10^{48}$~cm$^{-3}$) plotted against $T_{\rm GOES}$ (an isothermal plasma is assumed). Small circles (yellow) are means of the points over 1~MK intervals. The dotted and dashed curves are the theoretical $G(T_e)$ curves for the \ion{Ar}{17} $w$ line based on {\sc chianti} with the Ar abundance estimates of \cite{lod08} and \cite{asp09} respectively, and the ion fractions of \cite{bry09}. They include unresolved satellite lines.  (Right panel:) Histogram of argon abundances ($A({\rm Ar})$) in intervals of $A({\rm Ar}) = 0.1$, with the Ar abundances of \cite{lod08} and \cite{asp09} indicated as vertical lines (dotted and dashed respectively). The peak of the distribution gives $A({\rm Ar}) = 6.44$, and the distribution width (full width half maximum) defines the uncertainty range (6.35--6.52). Blue curves and vertical lines refer to the work of \cite{lod08}, red to the work of \cite{asp09}. } \label{G-of-T_ArXVII}
\end{figure}

To obtain estimates of the Ar abundance, we used a procedure similar to the one used by \cite{syl10} for the abundance of K from \ion{K}{18} lines observed in channel~1. {\it GOES} temperatures $T_{\rm GOES}$ and emission measures $EM_{\rm GOES}$ for the time interval corresponding to each spectrum were found and the value of the flux of each line divided by $EM_{\rm GOES}$ (units of $10^{48}$~cm$^{-3}$) was plotted against $T_{\rm GOES}$. The \ion{Ar}{17} $w$ line plot is shown in Figure~\ref{G-of-T_ArXVII} (left panel). Points from individual spectra are shown as black dots while averages over 1-MK temperature intervals are also shown. The much smaller scatter of points than for the corresponding plot for \ion{K}{19} lines (Fig. 3 of \cite{syl10}) reflects the larger fluxes of the \ion{Ar}{17} line. To these points are added theoretical contribution functions $G(T_e)$ given by

\begin{equation}
G(T_e) =  \frac{N({\rm Ar}^{+16}_i)}{N({\rm Ar}^{+16})} \frac{N({\rm Ar}^{+16})}{N({\rm Ar})} \frac{N({\rm Ar})}{N({\rm H})} \frac{N({\rm H})}{N_e} N_e A_{i0} \,\,\,\,\,\,{\rm cm}^3\,\,{\rm s}^{-1}
\end{equation}

\noindent where $N({\rm Ar}^{+16}_i) =$  population of the excited level $i = 1s2p\,^1P_1$ of ion Ar$^{+16}_i$, $N({\rm Ar}^{+16}) = $ number density of the ion Ar$^{+16}$ (all levels summed), $N({\rm Ar}) = $ the number density of all ionization stages of Ar, $N({\rm H}) = $ number density of hydrogen (H), $N_e$ is the electron density, and $A_{i0}$ is the transition probability from level $i$ to the ground state. Following \cite{phi08}, $N({\rm H})/N_e$ was taken to be 0.83, which is correct for coronal plasmas with $T_e > 10^5$~K. The atomic data, taken from version 6 of {\sc chianti}, include excitation rate coefficients from the Ar$^{+16}$ ground state, taken from \cite{whi01}, based on $R$-matrix code calculations of Ar$^{+16}$ up to $n=4$ levels including effects of radiation damping. \cite{whi01} estimate the uncertainties in the rate coefficients for excitation to the $1s2p\,^1P_1$ level to be $\pm 10$\%. The ion fractions $N({\rm Ar}^{+16})/N({\rm Ar})$ were taken from \cite{bry09}. We took the recent Ar abundance estimates $N({\rm Ar})/N({\rm H}) = A({\rm Ar})$ from both \cite{lod08} ($A({\rm Ar}) = 6.50$) and \cite{asp09} ($A({\rm Ar}) = 6.40$): different line-styles show the $G(T_e)$ curves for the two assumed abundances which are shown in Figure~\ref{G-of-T_ArXVII}. The contribution of unresolved \ion{Ar}{16} dielectronic satellites, which is accounted for in {\sc chianti}, is included in the $G(T_e)$ curves.

\begin{figure}
\epsscale{.80}
\plotone{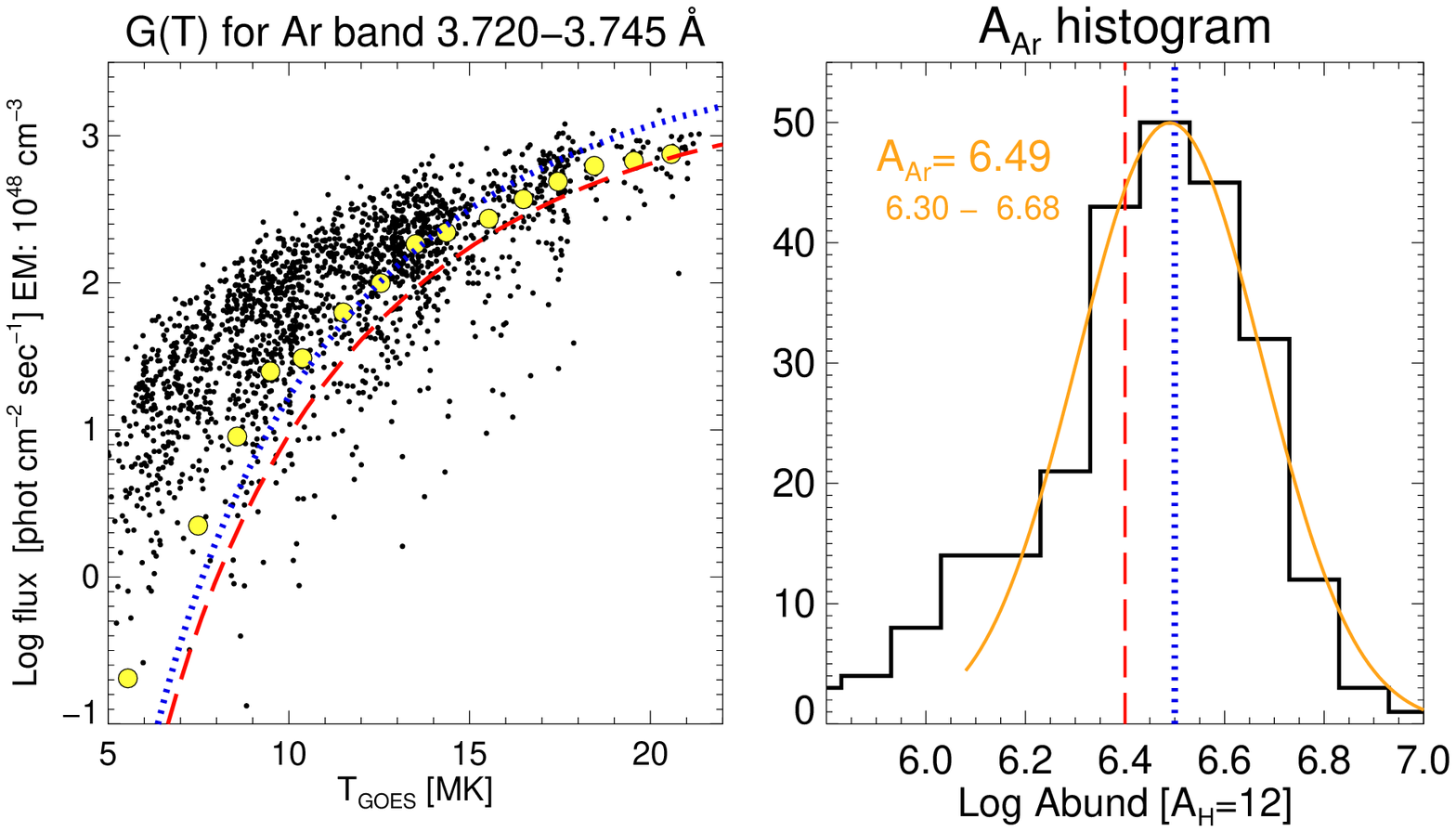}
\caption{(Left panel:) Estimated fluxes (points) in the \ion{Ar}{18} Ly-$\alpha$ line feature at 3.74~\AA\ divided by the {\it GOES} emission measure $EM_{\rm GOES}$ (units of $10^{48}$~cm$^{-3}$) plotted against $T_{\rm GOES}$. As with the analysis of the \ion{Ar}{17} lines, an isothermal plasma is assumed for all spectra. The theoretical curves are from {\sc chianti} with the Ar abundances of \cite{lod08} (dotted) and \cite{asp09} (dashed). (Right panel:) Histogram of argon abundances ($A({\rm Ar})$) plotted as in Figure~\ref{G-of-T_ArXVII}, for spectra having $T_{\rm GOES} > 15$~MK. Here the distribution gives $A({\rm Ar}) = 6.49$ with an uncertainty range from the distribution width (FWHM) of 6.30 to 6.68. Vertical lines indicate the argon abundance from \cite{lod08} (dotted) and from \cite{asp09} (dashed). Blue curves and vertical lines refer to the work of \cite{lod08}, red to \cite{asp09}.} \label{G-of-T_ArXVIII_Lya}
\end{figure}

A similar procedure was used for flux estimates of the \ion{Ar}{18} Ly-$\alpha$ line, where again {\it GOES} temperatures and emission measures were used. The resulting plot is shown in Figure~\ref{G-of-T_ArXVIII_Lya} (left panel). The theoretical contribution functions were also derived from {\sc chianti} and include unresolved satellite lines. In this case, excitation rate coefficients are taken from interpolation of $R$-matrix calculations of \cite{agg92} and \cite{agg93} for ionized He and Fe. This is a rather wide strength of atomic numbers, and there are now more recent data for specifically \ion{Ar}{18} lines \citep{agg08} but as the transitions are simple, it is thought that interpolation for the case of \ion{Ar}{18} ($Z=18$) will not lead to large errors.

Each value of the observed line flux divided by $EM_{\rm GOES}$ gives a value for the Ar abundance $N({\rm Ar})/N({\rm H})$, so a number distribution can be plotted for the abundance values obtained. This is done in the right-hand panels of Figures~\ref{G-of-T_ArXVII} and \ref{G-of-T_ArXVIII_Lya}. The derived value of $A({\rm Ar}) = 12 + {\rm log}_{10}\,[N({\rm Ar})/N({\rm H})]$ is plotted along the horizontal axis, and the numbers of spectra with particular values of $A({\rm Ar})$ over intervals of 0.1 are plotted along the vertical axis. An approximately gaussian distribution is apparent for the \ion{Ar}{17} $w$ line results, with the peak of the best-fit gaussian at $A({\rm Ar}) = 6.44$ and the FWHM points at 6.35 and 6.52. The value of $A({\rm Ar})$ from the \ion{Ar}{17} line fluxes is therefore $6.44 \pm 0.07$ (uncertainty is standard deviation). The \ion{Ar}{18} Ly-$\alpha$ line is formed at a higher temperature than the \ion{Ar}{17} $w$ line, and in fact the uncertainties in the wavelength interval defining the line (3.72--3.745~\AA) become larger than those in the nearby continuum for $T_{\rm GOES} < 15$~MK. We thus took only line flux measurements for $T_{\rm GOES} \gtrsim 15$~MK to derive the argon abundance. These estimates are shown in the right-hand panel of Figure~\ref{G-of-T_ArXVIII_Lya}, where as with the argon abundance estimates from the \ion{Ar}{17} $w$ line an approximately gaussian distribution is obtained. The  peak value $A({\rm Ar}) = 6.49$, with FWHM points at 6.30 and 6.68. The value of $A({\rm Ar})$  from the \ion{Ar}{18} line fluxes is therefore $6.49 \pm 0.16$. The larger uncertainty reflects the weaker flux of the \ion{Ar}{18} Ly-$\alpha$ and the smaller number of flux measurements. The mean of the \ion{Ar}{17} and \ion{Ar}{18} values weighted by the variances is $A({\rm Ar}) = 6.45 \pm 0.06$.

\begin{figure}
\epsscale{.80}
\plotone{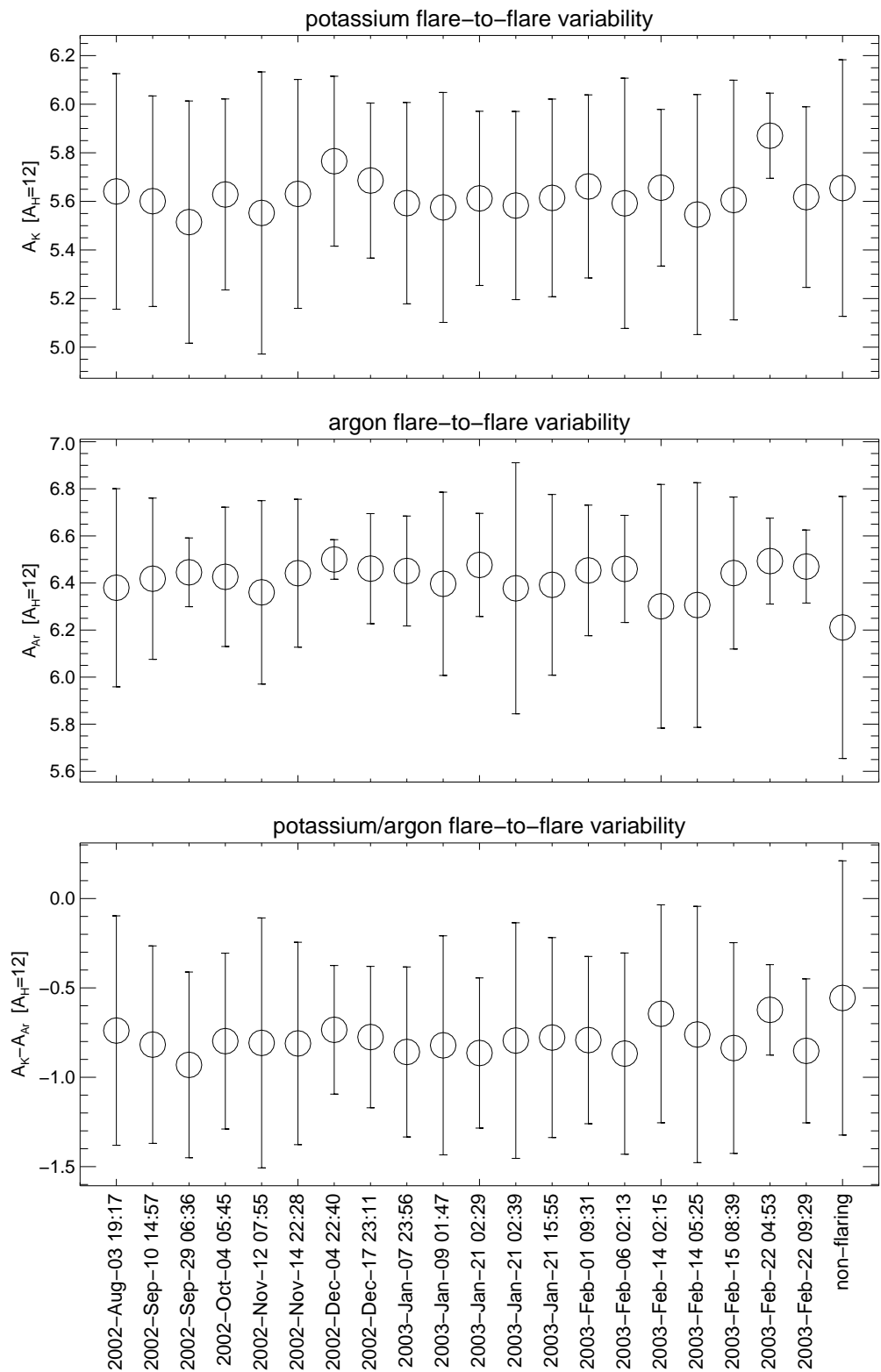}
\caption{Estimated abundances of K, $A({\rm K})$, from \ion{K}{18} line $w$ (3.53~\AA) and Ar, $A({\rm Ar})$, from \ion{Ar}{17} line $w$ (3.95~\AA), and the logarithm of the K/Ar abundance ratio, $A({\rm K}) - A({\rm Ar})$. For comparison an average of non-flare points is added (last point in each plot).} \label{flare-to-flare}
\end{figure}

There is apparently little variation from flare to flare in the derived values of $A({\rm Ar})$ from the \ion{Ar}{17} or \ion{Ar}{18} lines. This can be seen from Figure~\ref{flare-to-flare} (top panel) where the mean value of $A({\rm Ar})$ is plotted for each of the 20 flares here, together with a non-flaring period discussed in previous work \citep{bsyl10}. The value of $A({\rm K})$ is also plotted, being derived from the \ion{K}{19} lines in channel~1, as reported by \cite{syl10}, as well as the logarithm of the abundance ratio, i.e. $A({\rm K}) - A({\rm Ar})$. The error bars indicate the uncertainties in the mean values. The total variations are within the estimated uncertainties, and so the abundances of K and Ar appear to be constant for this sample of RESIK flares. The non-flaring points from \cite{bsyl10} are consistent with the remaining flare points though the uncertainties are somewhat larger because of the weaker \ion{K}{19} and \ion{Ar}{17} line emission at lower, non-flare temperatures. Possible time variations of the K and Ar abundances were checked because previous results from {\it Solar Maximum Mission} Bent Crystal Spectrometer data indicated time-varying abundances of Ca from the \ion{Ca}{19} $w$ line (3.177~\AA: see \cite{syl84}).

\section{DISCUSSION AND CONCLUSIONS}\label{Conclusions}

Our values of the Ar abundance $A({\rm Ar})$ from RESIK \ion{Ar}{17} $w$ line, $6.44 \pm 0.07$, and from the \ion{Ar}{18} Ly-$\alpha$ line, $6.49 \pm 0.16$, are very close to each other, the small difference being much less than the uncertainties indicated by the scatter of individual points in Figures~\ref{G-of-T_ArXVII} and \ref{G-of-T_ArXVIII_Lya}. The uncertainties in the excitation rate coefficients for the \ion{Ar}{17} lines are, as mentioned, $\pm 10$\%, which translates to a vertical shift of the theoretical contribution function $G(T_e)$ by this amount. This is very small compared with the scatter of measured values, however, even for the more intense \ion{Ar}{17} $w$ line. Even so, a slightly better estimate might result from use of the \ion{Ar}{17} $R$-matrix rate coefficients of \cite{agg05} and more particularly the \ion{Ar}{18} rate coefficients of \cite{agg08} since they are likely to be an improvement over the interpolated values presently used in the {\sc chianti} database for this ion.

Our abundance estimates may be compared with other published values. The estimate from RESIK spectra of \cite{phi03} is based on spectra from the 2002 period when the instrument characteristics were not quite optimized. In addition, a constant intensity calibration factor for each channel was taken though the more recent work of \cite{syl05} shows that there is a slight variation with wavelength. However, the value from \cite{phi03}, $A({\rm Ar}) = 6.45 \pm 0.03$, is very close to the value obtained here ($6.44 \pm 0.07$) from the \ion{Ar}{17} $w$ line.

Our values are somewhat less (up to 0.20 dex) than values obtained from  {\it Skylab} ultraviolet lines \citep{fel92a,fel00} but are very close to that obtained from {\it P78-1} observations of the \ion{Ar}{17} $1s^2-1s5p$ X-ray line at 3.128~\AA\ \citep{dos85}. The 3.128~\AA\ line used was not generally visible with the {\it P78-1} spectrometer but became apparent when the instrument was slightly off-pointed; the quality of the spectra are reasonably good and \cite{dos85} consider their estimate to be reliable. The {\it SMM} Flat Crystal Spectrometer scanned through the wavelength region of the \ion{Ar}{17} lines at 3.93~\AA\ during flares on a few occasions; the spectra, which do not have particularly good statistical quality, are discussed by \cite{phi94}. Observed ratios of the \ion{Ar}{17} $w$ line with the nearby \ion{S}{15} $1s^2-1s3p$ ($w3$) and $1s^2-1s4p$ ($w4$) lines gave a value of the Ar/S abundance ratio of $0.3 \pm 0.1$, which with the coronal value of $A({\rm S}) = 6.99$ from \cite{syl08b} leads to $A({\rm Ar}) = 6.47 \pm 0.15$, in agreement with the values obtained in this work.

The recent compilations of abundance data of \cite{lod08} and \cite{asp09} give $A({\rm Ar}) = 6.50 \pm 0.1$ and $A({\rm Ar}) = 6.40 \pm 0.13$ respectively, the chief difference being the abundance of O taken in these two works. Our value from \ion{Ar}{17} $w$ line flux measurements falls between these two values and agrees with both to within the stated uncertainties.

Some discussion on the nature of the FIP effect has resulted in different conclusions about coronal and proxies for photospheric argon abundances: \cite{fel92b} state that high-FIP (FIP $\gtrsim 10$~eV) elements have approximately the same abundances whereas \cite{flu99} argue that the coronal abundance of high-FIP elements are depleted by a factor 2. Considerably more data on argon abundances are now available since either of these publications, and are summarized by \cite{lod08}. A mean of her values of the proto-solar abundances, i.e. those assumed to be appropriate to the solar nebula, from a variety of objects (\ion{H}{2} regions, Jupiter's atmosphere, planetary nebulae, and B stars) is $A({\rm Ar}) = 6.55$ with a scatter of about $\pm 0.1$. Following \cite{lod08}, reducing this abundance by 0.07 to allow for heavy-element settling over the Sun's lifetime  gives a solar photospheric abundance of 6.48 which again is in agreement with our values. This suggests that there is little or no depletion of the coronal Ar abundance with respect to the photospheric, assuming that the adjusted proto-solar abundances are a reflection of the latter.

We acknowledge financial support from the European Commission's Seventh Framework Programme (FP7/2007--2013) under grant agreement No. 218816 (SOTERIA project, www.soteria-space.eu), the Polish Ministry of
Education and Science Grant N N203 381736,  and the UK--Royal Society/Polish Academy of Sciences International Joint Project (grant number 2006/R3) for travel support. {\sc chianti} is a collaborative project involving Naval Research Laboratory (USA), the Universities of Florence (Italy) and Cambridge (UK), and George Mason University (USA). We also thank J. Budaj, M. Dworetsky, K. Aggarwal, and F. P. Keenan for helpful advice in the analysis of the data.

\end{document}